\documentclass[12pt]{article}
\usepackage{amsfonts,amsmath}
\numberwithin{equation}{section}

\def\halft{{\textstyle\frac{1}{2}}}

\begin{document}

\title{The incompressible limit in linear anisotropic elasticity, with
application to surface waves and elastostatics}
\author{Michel Destrade, Paul A. Martin, Tom C.T. Ting}
\date{}

\maketitle

%
\begin{abstract}
Incompressibility is established for three-dimensional and two-di\-men\-sio\-nal deformations of an anisotropic linearly elastic material, as conditions to be satisfied by the elastic compliances.  These conditions make it straightforward to derive results for incompressible materials from those established for the compressible materials.  As an illustration, the explicit secular equation is obtained for surface waves in incompressible monoclinic materials with the symmetry plane at $x_3=0$.  This equation also covers the case of incompressible orthotropic materials. 
 
    The displacements and stresses for surface waves are often expressed in terms of the elastic stiffnesses, which can be unbounded in the incompressible limit.  An alternative formalism in terms of the elastic compliances presented recently by Ting is employed so that surface wave solutions in the incompressible limit can be obtained.  A different formalism, also by Ting, is employed to study the solutions to two-dimensional elastostatic problems.  
    
    In the special case of incompressible monoclinic material with the symmetry plane at $x_3=0$, one of the three Barnett-Lothe tensors $\mathbf{S}$ vanishes while the other two tensors $\mathbf{H}$ and $\mathbf{L}$ are the inverse of each other.  Moreover, $\mathbf{H}$ and $\mathbf{L}$ are diagonal with the first two diagonal elements being identical.  An interesting physical phenomenon deduced from this property is that there is no interpenetration of the interface crack surface in an incompressible bimaterial.  When only the inplane deformation is considered, it is shown that the image force due to a line dislocation in a half-space or in a bimaterial depends only on the magnitude, not on the direction, of the Burgers vector.
\end{abstract}

\newpage

\section{Introduction}

Linear anisotropic elasticity is characterized by two material constants, which can be taken as the shear modulus $\mu$  and Poisson's ratio $\nu$.  These constants satisfy $\mu>0$ and $-1<\nu<\halft$. The incompressible
limit is $\nu\to\halft$.  To see why this is so, we write down Hooke's law, relating the stress components  $\sigma_{ij}$ to the strain components $\epsilon_{kl}$ as
\begin{equation}
\sigma_{ij}=\mu\left( \frac{2\nu}{1-2\nu}\delta_{ij}\epsilon_{kk}
    +2\epsilon_{ij}\right).
\end{equation}
In the above,  is the Kronecker delta and repeated indices imply summation.  Contracting, we obtain
\begin{equation}
  \epsilon_{ii}= \frac{1-2\nu}{2\mu(1+\nu)}\sigma_{ii}
   =\frac{\nu}{\lambda(1+\nu)}\sigma_{ii},  \label{EPSIG}
\end{equation}
where $\lambda$ is a Lam\'e constant.  If the material is incompressible, $\epsilon_{ii}=0$ for every possible deformation, whence \eqref{EPSIG}$_1$ gives $\nu=\halft$.

Let us now turn to linear {\it anisotropic\/} elasticity, and consider the
corresponding incompressible limit. 
For such materials, we have
$\sigma_{ij} = C_{ijks} \epsilon_{ks}$, 
where the $C$'s are the elastic stiffnesses. 
In the special case of isotropy, the non-trivial stiffnesses are
$C_{1111} = C_{2222} = C_{3333}=\lambda +2 \mu$,
$C_{1122} = C_{1133}=C_{2233}=\lambda$  and
$C_{1212}=C_{1313}=C_{2323}=\mu$.
From \eqref{EPSIG}$_2$, the incompressible limit corresponds to
$\lambda\to\infty$. This suggests that, in general, some of the stiffnesses
will become unbounded in the incompressible limit, and therefore
it will be safer to
work with the coefficients of the elastic \textit{compliance}
matrix $ \mathbf{s}$ rather than with those of the elastic
stiffness matrix $ \mathbf{C}$. This is so because $ \mathbf{s}$
is the inverse of $ \mathbf{C}$, and possible
infinite components of $ \mathbf{C}$ will simply correspond to
some components (or combination of components) of $ \mathbf{s}$
being equal to zero.

In order to consider incompressible linearly elastic anisotropic materials directly, some authors have modified the stress-strain law by introducing a hydrostatic pressure $P$, as $\sigma_{ij} = - P \delta_{ij} + C_{ijks} \epsilon_{ks}$.  Incompressibility is then imposed by supplementing the condition $\epsilon_{ii}=0$.  Although formally acceptable, and supported by similar considerations in finite elasticity, this approach is risky as it may lead to potentially meaningless results, when the stiffnesses appear in their final expressions.

    For example, consider some recent developments in the theory of surface waves in linear anisotropic elastic materials.  For compressible materials the secular equation was obtained explicitly for monoclinic materials with the symmetry plane at $x_3=0$ 
\cite{Dest01a, Ting01}.  At the same time, some attention has been given to the consideration of interface waves in anisotropic materials which are \textit{incompressible} (see for instance \cite{NaSo99}
 or \cite{Dest01c}, and the references therein).  In this paper we show that results obtained in the general (compressible) case can be easily specialized to the incompressible case, simply by imposing the conditions for incompressibility on the elastic compliances, without having to introduce an arbitrary pressure.
    
    We adopt the following plan for the paper.  In Section 2 we recall the three-dimensional stress-strain laws of linear anisotropic elasticity, and establish that the constraint of incompressibility yields simple mathematical conditions, which are written for the elastic compliances $s_{\alpha \beta}$.  
Unlike the case of isotropic elastic materials, the conditions of incompressibility are different for \textit{two}-dimensional deformations.  
These conditions are established in Section 3 and written for the 
\textit{reduced elastic compliances} $s'_{\alpha \beta}$.  In both Sections, a necessary and sufficient condition for the strain energy density to be positive semidefinite is presented.  Then we show in Section 4 how simple it is to deduce an explicit secular equation for surface waves in a monoclinic material with the symmetry plane at $x_3 = 0$ for the incompressible case from that for the compressible case.  
The secular equation is only a part of the surface wave solution.  In the literature, the stresses and displacements for surface waves in an anisotropic elastic material are expressed in terms of the elastic stiffnesses, as briefly summarized in Section 5.  These expressions have to be converted to ones for the reduced elastic compliances.  
This has been done by Ting \cite{Ting01} and is outlined in Section 6.   The conversion presented in Section 6 does not apply to elastostatics.  A different formulation, again by Ting \cite{Ting99}, is reviewed in Section 7.  In Section 8 we consider the special case of incompressible monoclinic materials with the symmetry plane at $x_3=0$ under a static loading.  Interesting physical phenomena are discovered due to the incompressibility of the material.

\section{Incompressibility for three-dimensional deformations}

When the displacement $\mathbf{u}$ in an anisotropic linear elastic material 
depends on the three material coordinates $x_1$, $x_2$, $x_3$, 
the deformation is three-dimensional.  The relation between the strains $\epsilon_\alpha$ and the stresses $\sigma_\alpha$
in the contracted notation \cite{Voig10} is
 \begin{equation} \label{strain-stress3D}
  \epsilon_\alpha = s_{\alpha \beta}
  \sigma_\beta, 
\end{equation}
where $s_{\alpha \beta}$ are the elastic compliances. 
In particular, for isotropic materials, we have,
 \begin{equation} \nonumber
  \mathbf{s} =
   \frac{1}{2\mu(1+\nu)}
  \begin{bmatrix}
    1          &        &        &        &        &        \\
    -\nu       & 1      &        &        &        &        \\
    -\nu & -\nu & 1              &        &        &        \\
    0 & 0 & 0& 2(1+\nu)                   &        &        \\
    0 & 0 & 0 & 0 &      2(1+\nu) &                           \\
    0 & 0 & 0 & 0& 0 & 2(1+\nu)
  \end{bmatrix}.   \label{isocom}
\end{equation}

    In an \textit{incompressible} material the vanishing of the volume change is given by
\begin{equation}
  \epsilon_1 + \epsilon_2 + \epsilon_3
  = \sum_{\beta=1}^{6}\left( \sum_{\alpha=1}^{3} s_{\alpha \beta} \right)
   \sigma_\beta = 0.
\end{equation}
If this is to hold for any stresses we must have
\begin{equation}\label{incompr3D}
 \sum_{\alpha=1}^{3} s_{\alpha \beta} =0,
  \quad \text{ for } \beta = 1,2,3,4,5,6.
\end{equation}
There are six conditions for incompressibility.  
When the material is isotropic, \eqref{incompr3D} is trivially satisfied for $\beta = 4,5,6$ while for $\beta = 1,2,3$ 
it recovers the condition that $\nu = \halft$.

Now we show that \eqref{incompr3D} is \textit{structurally invariant} \cite{Ting01b}.  
If \eqref{incompr3D} holds for a coordinate system $x_j$, it holds for any other coordinate system $x^*_i$ 
obtained from $x_j$ by an orthogonal transformation $\mbox{\boldmath $\Omega$}$, say.  Let
\begin{equation} \label{rot}
 x^*_i = \Omega_{ij} x_j, \quad 
 \Omega_{ik} \Omega_{jk} = \delta_{ij} = \Omega_{ki}\Omega_{kj}
\end{equation}
In the four-index tensor notation, the elastic compliances $s_{ijks}$
referred to the rotated coordinate system $x^*_i$ become
\begin{equation}
 s^*_{ijks} = \Omega_{ip} \Omega_{jq} \Omega_{kr} \Omega_{st} s_{pqrt}.
\end{equation}
By contracting $i = j$ and using \eqref{rot}$_3$, this equation yields
\begin{equation} \label{rotContr}
 s^*_{iiks} =  \Omega_{kr} \Omega_{st} s_{pprt}.
\end{equation}
However, \eqref{incompr3D} in the four-index tensor notation is $s_{pprt}=0$.  
Equation \eqref{rotContr} then gives $s^*_{iiks}=0$.  This completes the proof.
    
The constraint \eqref{incompr3D} says that the first three rows of the 
$6 \times 6$ matrix $\mathbf{s}$ are linearly dependent.  This means that $\mathbf{s}$ is singular, 
and that the rank of $\mathbf{s}$ is at most five.  
We assume that the rank is five, because that is the case for isotropic materials.  The strain energy density cannot be negative for an incompressible material.  
Hence $\mathbf{s}$ must be positive semidefinite.  The rank of $\mathbf{s}$ being five implies that there exists a 
$5 \times 5$ submatrix that is non-singular.  According to a theorem presented in \cite{Hohn65}, a necessary and sufficient condition for the matrix $\mathbf{s}$ of rank five to be positive semidefinite is that the five leading principal minors of the non-singular submatrix be positive.  It means that this non-singular submatrix must be positive definite.

To apply the theorem we write the matrix $\mathbf{s}$ 
satisfying the constraint \eqref{incompr3D} in the form
\begin{equation} \label{s6x6}
  \mathbf{s} =
  \begin{bmatrix}
    s_{22} + 2s_{23} + s_{33} 
                     &        &        &        &        &       \\
    -(s_{22} + s_{23})
                     & s_{22} &        &        &        &       \\
    -(s_{23} + s_{33})
                    & s_{23} & s_{33} &        &        &        \\
    -(s_{24} + s_{34})
                    & s_{24} & s_{34} & s_{44} &        &        \\
    -(s_{25} + s_{35})
                    & s_{25} & s_{35} & s_{45} & s_{55} &       \\
    -(s_{26} + s_{36})
                    & s_{26} & s_{36} & s_{46} & s_{56} & s_{66}
  \end{bmatrix}.
\end{equation}
Only the lower triangle of the matrix is shown since it is symmetric.  
The $5 \times 5$ submatrix on the lower right corner of $\mathbf{s}$ can be prescribed arbitrarily and the elements in the first column (and hence the first row) of $\mathbf{s}$ are then determined.  
We will therefore take the $5 \times 5$ submatrix on the lower right corner of $\mathbf{s}$ to be non-singular.  
Before we write down the leading principal minors of this submatrix, we introduce the following notation for the minors of $\mathbf{s}$ .  
Let $s(n_1, \ldots, n_k | m_1, \ldots, m_k)$ be the $k \times k$ 
minor of the matrix $s_{\alpha \beta}$, 
the elements of which belong to the rows of $s_{\alpha \beta}$ numbered $n_1, \ldots n_k$
and columns numbered $m_1, \ldots m_k$, $1 \le k \le 6$.  
A principal minor is $s(n_1, \ldots, n_k | n_1, \ldots, n_k)$, which is written as $s(n_1, \ldots, n_k)$ for simplicity.  
If the leading principal minors are taken from the lower right corner of the submatrix, a necessary and sufficient condition for the matrix 
$\mathbf{s}$ to be positive semidefinite is
\begin{equation}\label{iff1}
  s_{66} > 0, \quad s(5,6) > 0, \quad s(4,5,6) > 0, 
  \quad s(3,4,5,6) > 0, \quad s(2,3,4,5,6) > 0.
\end{equation}
If they are taken from the top left corner of the submatrix, we have
 \begin{equation}\label{iff2}
  s_{22} > 0, \quad s(2,3) > 0, \quad s(2,3,4) > 0, 
  \quad s(2,3,4,5) > 0, \quad s(2,3,4,5,6) > 0.
\end{equation}
Equation \eqref{iff1} or \eqref{iff2} 
is the necessary and sufficient condition for the matrix $\mathbf{s}$ to be positive semidefinite.

The first two inequalities in \eqref{iff2} 
are the necessary and sufficient conditions for the $3 \times 3$ submatrix on the top left corner of the matrix 
$\mathbf{s}$ to be positive semidefinite.  When the three equations for $\beta = 1,2,3$ in \eqref{incompr3D} 
are solved for $s_{12}, s_{23}, s_{31}$, we have
\begin{equation} \label{s12s23s31}
s_{12} = \halft (s_{33} - s_{11} - s_{22}), \quad 
s_{23} = \halft (s_{11} - s_{22} - s_{33}), \quad 
s_{31} = \halft (s_{22} - s_{33} - s_{11}). 
\end{equation}
Hence $s_{11}, s_{22}, s_{33}$ are all we need to prescribe the $3 \times 3$ submatrix.  
The $s_{11}, s_{22}, s_{33}$  are, respectively, $1/E_1, 1/E_2, 1/E_3$, where $E_i$ are the Young's moduli.  
With the $s_{23}$ given in \eqref{s12s23s31}, the second inequality in \eqref{iff2} is
\begin{equation}\label{2ndIneq}
s(2,3) - s_{22} s_{33} 
- \textstyle{\frac{1}{4}} (s_{11} - s_{22} - s_{33})^2>0.
\end{equation}
Since $s_{22}>0$, equation \eqref{2ndIneq} tells us that $s_{33}>0$.  
Equation \eqref{2ndIneq} can then be written as
\begin{equation}
[(\sqrt{s_{22}} + \sqrt{s_{33}})^2 - s_{11}]
[s_{11} - (\sqrt{s_{22}} - \sqrt{s_{33}})^2] > 0.
\end{equation}
It tells us that $s_{11}>0$.  
This is rewritten in a form symmetric with respect to $s_{11}, s_{22}, s_{33}$ 
as
\begin{equation}\label{symm}
(U+V+W)(U+V-W)(V+W-U)(W+U-V)>0,
\end{equation}
where $U = \sqrt{s_{11}}, V = \sqrt{s_{22}}, W = \sqrt{s_{33}}$.  
Scott \cite{Scot00} obtained the same inequality, involving the area modulus of elasticity.  From Hero's formula, the left hand side of \eqref{symm} is, 
after taking the square root and dividing the result by 4, the area of a triangle whose three sides are $U, V, W$.  Thus $\sqrt{s_{11}}, \sqrt{s_{22}}, \sqrt{s_{33}}$ 
must form a triangle with a nonzero area for the $3 \times 3$ submatrix to be positive semidefinite.

Another geometrical interpretation of the constraint on $s_{11}, s_{22}, s_{33}$
can be made by noticing that \eqref{symm} is equivalent to
\begin{equation}
 V+W > U > |V-W|.
\end{equation}
In a rectangular coordinate system $U, V, W$, the point ($U, V, W$) 
is inside a triangular cone (or pyramid) in the space $U>0, V>0, W>0$.  
The three edges of the cone lie on the three coordinate planes making an equal angle ($\pi/4$) with the coordinate axes.

When the material is compressible, Zheng and Chen \cite{ZhCh00} employed the notation
\begin{equation}
 n_i = \frac{-s_{jk}}{s_{jj}s_{kk}} 
  = \sqrt{\frac{E_k}{E_j}} \nu_{jk}
  = \sqrt{\frac{E_j}{E_k}} \nu_{kj},
\end{equation}
where $\nu_{ij}$ are Poisson's ratios and $\{i, j, k\}$ 
is a cyclic permutation of $\{1,2,3\}$.  The condition for the $3 \times 3$ submatrix to be positive definite is $|n_i| <1$, 
($i=1,2,3$) and
 \begin{equation} \label{zhongi}
 n_1^2 + n_2^2 + n_3^2 + 2 n_1 n_2 n_3 <1.
\end{equation}
The geometry of the solid represented by \eqref{zhongi} 
resembles that of a Chinese delicacy called \textit{Zongzi}.  
For an incompressible material, ($n_1, n_2, n_3$) lies on the surface of 
a \textit{Zongzi}.    

Equation \eqref{2ndIneq} can be written in a symmetric form as
\begin{equation}
  s(2,3) = \halft (s_{11}s_{22} + s_{22}s_{33} + s_{33}s_{11})
   -  \textstyle{\frac{1}{4}}  (s_{11} + s_{22} + s_{33})^2 >0.
\end{equation}
Hence the three $2 \times 2$ minors $s(2,3), s(3,1), s(1,2)$ are identical.

\section{Incompressibility for two-dimensional deformations}

When the displacement $\mathbf{u}$ depends on $x_1, x_2$,  but not on $x_3$, 
the deformation is two-dimensional.  In this case $ \epsilon_3 = u_{3,3} = 0$, and \eqref{strain-stress3D} is replaced by
\begin{equation} \label{strain-stress2D}
\epsilon_\alpha = s'_{\alpha \beta} \sigma_\beta,
\end{equation}
where
\begin{equation} \label{reduced}
  s'_{\alpha \beta}
   = s_{\alpha \beta}
     - \frac{ s_{\alpha 3} s_{3 \beta}}{s_{33}},
\end{equation}
are the reduced elastic compliances \cite{Lekh63}. 
It should be noted that $s'_{\alpha 3} = s'_{3 \alpha} = 0$. With
\eqref{strain-stress2D}, the incompressibility condition
$\epsilon_1 + \epsilon_2 = 0$ yields
\begin{equation}\label{incompr2D}
  s'_{1 \beta} + s'_{2 \beta} =0,
  \quad \text{ for } \beta = 1,2,4,5,6.
\end{equation}
When the material is isotropic, \eqref{incompr2D} is trivially satisfied 
for $\beta =4,5,6$, while for $\beta = 1,2$, it recovers the condition that $\nu = \halft$.

Under a rotation of the coordinate system about the $x_3$-axis, 
Ting \cite{Ting01b} has shown that the following relations for the elastic stiffnesses $C_{\alpha \beta}$ 
in the contracted notation are \textit{structurally invariant}:
\begin{equation} \label{strucInv1}
  C_{16} + C_{26} = C_{11} - C_{22} =0, \quad
  C_{14} + C_{24} = C_{15} + C_{25} =0.
\end{equation}
They are called Type 1A and 4A, respectively.  
He pointed out that \eqref{strucInv1} applies also to $s'_{\alpha \beta}$.  Following his derivation it can be shown that
\begin{equation} \label{strucInv2}
  C_{11} + C_{12} = C_{12} + C_{22} 
  = C_{16} + C_{26} = 0
\end{equation}
is structurally invariant, and that it applies to $s'_{\alpha \beta}$.  
Thus the incompressibility condition \eqref{incompr2D} is \textit{structurally invariant} 
under rotation of the coordinate axes about the $x_3$-axis.

The reduced elastic compliance matrix that satisfies 
\eqref{incompr2D} has the structure
\begin{equation}
 \mathbf{s'} =
 \begin{bmatrix}
    s'_{22} &         &         &         &          \\
   -s'_{22} & s'_{22} &         &         &          \\
   -s'_{24} & s'_{24} & s'_{44} &         &          \\
   -s'_{25} & s'_{25} & s'_{45} & s'_{55} &          \\
   -s'_{26} & s'_{26} & s'_{46} & s'_{56} &  s'_{66} \
  \end{bmatrix}.
\end{equation}
The matrix $\mathbf{s'}$ must be positive semidefinite. A necessary and sufficient condition for the matrix $\mathbf{s'}$ to be positive semidefinite is that the four leading principal minors of the $4 \times 4$ submatrix on the lower right corner of $\mathbf{s'}$ be positive.  If the leading principal minors are taken from the lower right corner of the submatrix, a necessary and sufficient condition for $\mathbf{s'}$ to be positive semidefinite is
\begin{equation}
  s'_{66} > 0, \quad s'(5,6) > 0, \quad s'(4,5,6) > 0, 
  \quad s'(2,4,5,6) > 0.
\end{equation}
If they are taken from the top left corner of the submatrix, we have
\begin{equation}
  s'_{22} > 0, \quad s'(2,4) > 0, \quad s'(2,4,5) > 0, 
  \quad s'(2,4,5,6) > 0.
\end{equation}

Using \eqref{reduced}, equation \eqref{incompr2D} can be rewritten as
\begin{equation}
s_{1\beta} + s_{2 \beta} + w s_{3 \beta} = 0,
\quad w = -(s_{13} + s_{23})/s_{33}.
\end{equation}
It is an identity when $\beta=3$.  
An elastic compliance matrix that satisfies this equation has the structure
\begin{equation} \label{s5x5}
  \mathbf{s} =
  \begin{bmatrix}
    s_{22} + 2s_{23} + w^2 s_{33} 
                     &        &        &        &        &       \\
    -(s_{22} + w s_{23})
                     & s_{22} &        &        &        &       \\
    -(s_{23} + w s_{33})
                    & s_{23} & s_{33} &        &        &        \\
    -(s_{24} + w s_{34})
                    & s_{24} & s_{34} & s_{44} &        &        \\
    -(s_{25} + w s_{35})
                    & s_{25} & s_{35} & s_{45} & s_{55} &       \\
    -(s_{26} + w s_{36})
                    & s_{26} & s_{36} & s_{46} & s_{56} & s_{66}
  \end{bmatrix}.
\end{equation}
where $w$ is arbitrary.  It reduces to \eqref{s6x6} when $w=1$.  
Thus incompressibility in three-dimensional deformations implies incompressibility in two-dimensional deformations, but the converse need not hold.  A necessary and sufficient condition for the matrix $\mathbf{s}$ in \eqref{s5x5} to be positive semidefinite is identical to the one in \eqref{iff1} or \eqref{iff2}.  It should be noted that \eqref{iff1} or \eqref{iff2} does not involve $w$.  If the matrix $\mathbf{s}$ in \eqref{s5x5} is positive semidefinite for any $w$, then $s_{11}$ and $s(1,2)$, which can be computed easily, should be non-negative for any $w$.  It can be shown that
\begin{equation}
s_{11} = (w \sqrt{s_{33}} + \frac{s_{23}}{\sqrt{s_{33}}})^2 + \frac{s(2,3)}{s_{33}},
\quad
s(1,2) = w^2 s(2,3),
\end{equation}
so that $s_{11}$ and $s(1,2)$ are indeed non-negative for any $w$.  
When $w = 0$, $s(1,2) = 0$ but the rank of the $3 \times 3$ submatrix on the top left corner of the matrix $\mathbf{s}$ is two for any $w$.

\section{Secular equation for surface waves in incompressible monoclinic materials}

The interest for considering incompressibility for surface waves in linear anisotropic elasticity is threefold.  From a \textit{historical} perspective, it must be remembered that Rayleigh, the initiator of the theoretical study of elastic surface waves, did treat the case of an incompressible linearly isotropic elastic half-space \cite{Rayl85}.  Although some literature can be found on the subject of surface waves in incompressible, finitely elastic, stress-induced anisotropic half-spaces \cite{Flav63, Will73, DoOg90, Chad97}, very few papers are placed within the counterpart context of linearly elastic, anisotropic half-spaces, subject to the internal constraint of incompressibility.  Second, from an \textit{experimental} point of view, it is accepted \cite{NaSo99,NaSo97,SoNa99,Sutc92,GuGu99} that certain elastic materials may be modeled as incompressible, linearly elastic, anisotropic materials.  According to Nair and Sotiropoulos \cite{NaSo97}, such is the case for ``polymer Kratons, thermoplastic elastomers, rubber composites when low frequency waves are considered to justify the assumption of material inhomogeneity, etc''.  Third, the \textit{theoretical} aspect of incompressibility in linear anisotropic elasticity has not been addressed in this context, and it is important to derive the secular equation in terms of the compliances rather than in terms of the stiffnesses.

Here attention is turned to surface waves propagating with speed $v$ in the direction of the $x_1$-axis in the half-space $x_2>0$.  The material is monoclinic with the symmetry plane at $x_3=0$.  In the general (compressible) case the secular equation for the surface wave has been obtained explicitly by Destrade \cite{Dest01a} using the method of first integrals introduced by Mozhaev \cite{Mozh95}, and by Ting \cite{Ting01} using a modified Stroh \cite{Stro62} formalism.  Letting $X = \rho v^2$ where $\rho$ is the mass density, the secular equation is
\begin{multline} \label{secularCompr}
[\eta - (1+r_2)X]
 \{(\eta - X) [ (\eta - X)(n_{66}X -1) + r_6^2 X]
     + X^2[ (\eta - X)n_{22} + r_2^2] \} \\
 + 2 r_6 X^2 (\eta - X) [(\eta - X) n_{26} + r_2 r_6]
  = 0.
\end{multline}
It is a quartic in X.  In \eqref{secularCompr}, (see \cite{Ting01})
\begin{align} \label{constants}
& \eta = \frac{1}{s'_{11}},
    \quad r_2 = -\frac{s'_{12}}{s'_{11}},
        \quad r_6 = -\frac{s'_{16}}{s'_{11}},  \nonumber \\
& n_{66} = \frac{s'(1,6)}{s'_{11}}, \quad
    n_{26} = \frac{s'(1,2|1,6)}{s'_{11}}, \quad
        n_{22} = \frac{s'(1,2)}{s'_{11}}.
\end{align}

The incompressible case was first studied by Nair and Sotiropoulos \cite{NaSo99}, although they did not establish the secular equation explicitly.  The secular equation for incompressible materials can be deduced directly from \eqref{secularCompr} by imposing the incompressibility conditions $s'_{2 \beta} = -s'_{1 \beta}$.  The $r_2, n_{26}, n_{22}$ in \eqref{constants} simplify to
\begin{equation}
r_2 =1, \quad n_{26} = 0, \quad n_{22} = 0,
\end{equation}
and the secular equation \eqref{secularCompr} reduces to
\begin{equation}
(\eta - 2X) [(\eta - X)^2 (n_{66}X -1) + X^2]
 +  r_6^2 \eta X(\eta - X)  = 0.
\end{equation}
It can be written in a non-dimensional form as
\begin{align}
& (1- 2 \xi)[(1-\xi)^2 (\kappa \xi - 1) + \xi^2] + r_6^2 \xi (1 - \xi)
=  0, \\
& \xi = X/\eta = \rho v^2/s'_{11}, \quad 
\kappa =n_{66}/s'_{11}.
\end{align}

For incompressible orthotropic materials for which $s'_{16}=0$, the secular equation further simplifies to, since $(1- 2 \xi) \ne 0$,
\begin{equation}
(1-\xi)^2 (1 - \kappa \xi) = \xi^2, \quad 
\kappa = s'_{66}/s'_{11}.
\end{equation}
This cubic in $\xi$ has a more compact and satisfying form than that obtained in terms of the stiffnesses \cite{Dest01c} which, as stressed in the Introduction, are not easily defined for incompressible anisotropic materials.
    The secular equation is only a part of the surface wave solution.  A complete solution requires the computation of the displacements and stresses.  This is discussed next.

\section{The Stroh formalism for steady state motion}

In a fixed rectangular coordinate system $x_i$ ($i=1,2,3$) the stress-strain law and the equations of motion are
\begin{align}
\label{stress-strain}
& \sigma_{ij} = C_{ijks} u_{k,s}, \\
\label{motion}
& C_{ijks} u_{k,sj} = \rho \ddot{u}_i.
\end{align}
in which the dot stands for differentiation with time $t$.  Consider a steady state motion with the steady wave speed $v$ propagating in the direction of the $x_1$-axis.  A solution for the displacement vector $\mathbf{u}$ of \eqref{motion} can be written as \cite{Stro62}
\begin{equation} \label{displVector}
\mathbf{u} = \mathbf{a} f(z), \quad z = x_1 - vt + p x_2,
\end{equation}
in which $f$ is an arbitrary function of $z$, and $p$ and  $\mathbf{a}$  satisfy the eigenrelation
\begin{align}
\label{eigen1}
& \mbox{\boldmath $\Gamma$} \mathbf{a} = \mathbf{0}, \\
& \mbox{\boldmath $\Gamma$} = \mathbf{Q} - X \mathbf{I}
  + p( \mathbf{R} + \mathbf{R}^T) + p^2 \mathbf{T}, \\
& X = \rho v^2.
\end{align}
In the above the superscript $T$ stands for the transpose, $ \mathbf{I}$ is the unit matrix, and $ \mathbf{Q}, \mathbf{R}, \mathbf{T}$ are $3 \times 3$ matrices whose elements are
\begin{equation}
Q_{ik} = C_{i1k1} , \quad
R_{ik} = C_{i1k2} , \quad
T_{ik} = C_{i2k2}.
\end{equation}
The matrices $\mathbf{Q}$ and $\mathbf{T}$ are symmetric and so is the matrix $\mbox{\boldmath $\Gamma$}$.  Introducing the new vector $ \mathbf{b}$ defined by
\begin{equation} \label{b}
\mathbf{b} = ( \mathbf{R}^T + p \mathbf{T}) \mathbf{a}
 = -[ p^{-1} (\mathbf{Q} - X \mathbf{I}) + \mathbf{R}] \mathbf{a},
\end{equation}
in which the second equality follows from \eqref{eigen1}, the stress determined from 
\eqref{stress-strain} can be written as
\begin{equation}
\sigma_{i1} = - \phi_{i,2} - \rho v \dot{u}_i, \quad
\sigma_{i2} =  \phi_{i,1}.
\end{equation}
The $\phi_i$ ($i=1,2,3$) are the components of the stress function vector
\begin{equation} \label{stressVector}
\mbox{\boldmath $\phi$} = \mathbf{b} f(z).
\end{equation}

There are six eigenvalues $p_\alpha$ and six Stroh eigenvectors $ \mathbf{a}_\alpha$ and  $ \mathbf{b}_\alpha$($\alpha = 1,2,\ldots,6$).  When $p_\alpha$ are complex, they consist of three pairs of complex conjugates.  If $p_1, p_2, p_3$ are the eigenvalues with a positive imaginary part, the remaining three eigenvalues are the complex conjugates of $p_1, p_2, p_3$.  Assuming that $p_1, p_2, p_3$ are distinct, the general solution obtained from superposing three solutions of \eqref{displVector} and \eqref{stressVector} associated with $p_1, p_2, p_3$  can be written in matrix notation as
\begin{equation} \label{u-phi}
\mathbf{u} = \mathbf{A} <f(z_*)> \mathbf{q},
\quad
\mbox{\boldmath $\phi$} = \mathbf{B} <f(z_*)> \mathbf{q},
\end{equation}
where $ \mathbf{q}$ is an arbitrary constant vector and
\begin{align} 
& \mathbf{A} = [ \mathbf{a}_1, \mathbf{a}_2, \mathbf{a}_3], \quad
 \mathbf{B} = [ \mathbf{b}_1, \mathbf{b}_2, \mathbf{b}_3],
 \nonumber \\
& <f(z_*)> = \text{Diag} [ f(z_1), f(z_2), f(z_3)], \label{AB}\\
&  z_\alpha = x_1 - vt + p_\alpha x_2. \nonumber
\end{align}

For surface waves in the half-space $x_2 \ge 0$, the function $f(z)$ is chosen as
\begin{equation}
f(z) = e^{ikz},
\end{equation}
where $k$ is the real wave number.  Since the imaginary parts of $p_1, p_2, p_3$ are positive, \eqref{u-phi}$_1$ assures us that $\mathbf{u} \to \mathbf{0}$ as $x_2 \to \infty$.  The surface traction at $x_2=0$ vanishes if 
$\mbox{\boldmath $\phi$} = \mathbf{0}$ at $x_2=0$, i.e.,
\begin{equation} 
\mathbf{B} \mathbf{q} = \mathbf{0}.
\end{equation}
This has a nontrivial solution for $\mathbf{q}$ when the determinant of $\mathbf{B}$ vanishes, i.e.,
\begin{equation} \label{B=0}
| \mathbf{B} | = 0.
\end{equation}
This is the secular equation for $v$.  For a monoclinic material with the symmetry plane at $x_3=0$,  \eqref{B=0} leads to \eqref{secularCompr}.

The displacement $\mathbf{u}$ and the stress function vector $\mbox{\boldmath $\phi$}$ given in \eqref{u-phi} require the computation of the eigenvalues $p_\alpha$ and the eigenvectors  $ \mathbf{a}_\alpha$ and $ \mathbf{b}_\alpha$ ($\alpha=1,2,3$).  They are provided by \eqref{eigen1} and \eqref{b} which are in terms of the elastic stiffnesses.  They are not suitable for taking the incompressible limit.  A different expression in terms of the reduced elastic compliances is needed.  This is presented next.

\section{Steady state motion for incompressible materials}

The two equations in \eqref{b} can be written in a standard eigenrelation as \cite{InTo69, BaLo73, ChSm77}
\begin{align}  \label{Nxi}
& \mathbf{N}\mbox{\boldmath $\xi$}  = p \mbox{\boldmath $\xi$}, \\
& \mathbf{N} =  
  \begin{bmatrix}
     \mathbf{N}_1                 &   \mathbf{N}_2    \\
     \mathbf{N}_3 +  X \mathbf{I} &   \mathbf{N}_1^T  
  \end{bmatrix}, \quad  
    \mbox{\boldmath $\xi$}  =    
       \begin{bmatrix}
           \mathbf{a}  \\
           \mathbf{b}  
        \end{bmatrix},    \\
& \mathbf{N}_1 = -\mathbf{T}^{-1} \mathbf{R}^T, \quad
  \mathbf{N}_2 = \mathbf{T}^{-1}, \quad
  \mathbf{N}_3 =  \mathbf{R}\mathbf{T}^{-1}\mathbf{R}^T - \mathbf{Q}.
\end{align}
It was shown in \cite{Ting88} that $ \mathbf{N}_1, \mathbf{N}_2, \mathbf{N}_3$ have the structure
\begin{equation}
 \label{eigen2}
 -\mathbf{N}_1 =  
   \begin{bmatrix}
     r_6     &   1    &  s_6    \\
     r_2     &   0    &  s_2    \\ 
     r_4     &   0    &  s_4 
  \end{bmatrix}, 
 \mathbf{N}_2 =  
   \begin{bmatrix}
     n_{66}  &   n_{26}  &  n_{46}    \\
     n_{26}  &   n_{22}  &  n_{24}    \\ 
     n_{46}  &   n_{24}  &  n_{44}    
   \end{bmatrix},  
 -\mathbf{N}_3 =  
   \begin{bmatrix}
     m_{55}  &   0  &  -m_{15}    \\
     0       &   0  &  0          \\ 
    -m_{15}  &   0  &  m_{11}    
   \end{bmatrix}.   
\end{equation}
An explicit expression of the elements of $ \mathbf{N}_1, \mathbf{N}_2, \mathbf{N}_3$ was given in \cite{ChSm77} (see also \cite[p. 167]{Ting96} in terms of the reduced elastic compliances and in \cite{BaCh90} in terms of the elastic stiffnesses.  The expressions in term of the reduced elastic compliances are
\begin{align}
& r_\alpha = \frac{1}{\Delta} s'(1,5|5,\alpha), \quad
    s_\alpha = \frac{1}{\Delta} s'(1,5|\alpha,1),  \nonumber \\
& n_{\alpha \beta} = \frac{1}{\Delta} s'(\alpha,1,5|\beta,1,5),
\quad m_{\alpha \beta} = \frac{1}{\Delta} s'_{\alpha \beta},
\quad  \Delta = s'(1,5).
\end{align}

Since $s'_{2\beta} = - s'_{1\beta}$ for incompressible materials, it can be shown that
\begin{equation}
r_2 =1, \quad s_2 =0, \quad n_{26}=n_{22}=n_{24}=0.
\end{equation}
Thus, for incompressible materials, the matrices $ \mathbf{N}_1$ and $\mathbf{N}_2$ have the simpler expressions (see also Chadwick \cite{Chad97}),
\begin{equation}
 -\mathbf{N}_1 =  
   \begin{bmatrix}
     r_6     &   1    &  s_6    \\
     1       &   0    &  0      \\ 
     r_4     &   0    &  s_4 
  \end{bmatrix}, 
 \mathbf{N}_2 =  
   \begin{bmatrix}
     n_{66}  &   0  &  n_{46}    \\
     0       &   0  &  0         \\ 
     n_{46}  &   0  &  n_{44}    
   \end{bmatrix}.   
\end{equation}
Equation \eqref{Nxi} consists of six scalar equations.  The second and the fifth equations provide the identities
\begin{equation} \label{a1a2b1b2}
a_1 + p a_2 =0, \quad b_1 + p b_2 = X a_2.
\end{equation}
The first identity could have been deduced by inserting the solution \eqref{displVector} into the condition of incompressibility
\begin{equation} \label{incompCond}
\epsilon_1 + \epsilon_2 = u_{1,1} + u_{2,2} = 0.
\end{equation}
With $ \mathbf{N}_1, \mathbf{N}_2, \mathbf{N}_3$ expressed in terms of $s'_{\alpha\beta}$, equation \eqref{Nxi} can be employed to compute the eigenvalues $p$ and the eigenvectors $\mathbf{a}$ and $\mathbf{b}$.  Equation \eqref{Nxi} consists of two equations,
\begin{equation} \label{eigen3}
(\mathbf{N}_1 - p \mathbf{I}) \mathbf{a} + \mathbf{N}_2 \mathbf{b} = \mathbf{0},
\quad
(\mathbf{N}_3 + X\mathbf{I}) \mathbf{a} + (\mathbf{N}_1^T - p \mathbf{I})\mathbf{b} = \mathbf{0}.
\end{equation}
Assuming that $(\mathbf{N}_3 + X\mathbf{I})$ in not singular, \eqref{eigen3}$_2$ can be solved for $\mathbf{a}$ and \eqref{eigen3}$_1$ can be written as \cite{Ting01}
\begin{align}
\label{eigen4}
& \widehat{\mbox{\boldmath $\Gamma$}} \mathbf{b} = \mathbf{0}, \\
& \widehat{\mbox{\boldmath $\Gamma$}} = \widehat{\mathbf{Q}}
  + p( \widehat{\mathbf{R}} + \widehat{\mathbf{R}}^T) + p^2 \widehat{\mathbf{T}}.
\end{align}
In the above,
\begin{equation} \label{ThatRhatQhat}
\widehat{\mathbf{T}}= (\mathbf{N}_3 + X\mathbf{I})^{-1},
\quad
\widehat{\mathbf{R}} = \mathbf{N}_1 \widehat{\mathbf{T}},
\quad
\widehat{\mathbf{Q}} = \mathbf{N}_1 \widehat{\mathbf{T}}\mathbf{N}_1^T + \mathbf{N}_2.
\end{equation}
The matrices $\widehat{\mathbf{T}}$ and $\widehat{\mathbf{Q}}$ are symmetric, so is 
$\widehat{\mbox{\boldmath $\Gamma$}}$.  Equation \eqref{eigen4} provides the eigenvalue $p$ and the eigenvector $\mathbf{b}$.  The eigenvector $\mathbf{a}$ obtained from \eqref{eigen3}$_2$ is, using \eqref{eigen4} and \eqref{ThatRhatQhat},
\begin{equation} \label{a}
\mathbf{a} = -(\widehat{\mathbf{R}}^T +p \widehat{\mathbf{T}})\mathbf{b}
= (p^{-1} \widehat{\mathbf{Q}} + \widehat{\mathbf{R}})\mathbf{b}.
\end{equation}
We have thus presented equations for computing the eigenvalues $p$ and the eigenvectors $ \mathbf{a}$ and $\mathbf{b}$ needed for the surface wave solution in terms of $s'_{\alpha\beta}$.  The surface wave solution for an incompressible material is then complete.

\section{Elastostatics for incompressible materials}
 
The solutions \eqref{u-phi} and \eqref{AB} remain valid for elastostatics if we set $v=0$.  The derivation in \eqref{Nxi}-\eqref{incompCond} also holds for elastostatics if we let $X=0$.  However, the derivation from \eqref{eigen4} to \eqref{a} is not valid for elastostatics because $(\mathbf{N}_3 + X\mathbf{I})$ is singular when $X=0$.  
A different approach is needed to find $ \mathbf{a}$ and $\mathbf{b}$ in terms of $s'_{\alpha\beta}$.

A modified Lekhnitskii formalism in the style of Stroh was proposed by Ting \cite{Ting99} in which the vector $\mathbf{b}$ satisfies the eigenrelation (see also \cite{BaCh90, BaKi97})
\begin{equation} \label{eigenStatics}
    \begin{bmatrix}
     1     &   -p    &  0    \\
     0     &   l_4   & -l_3 \\ 
     0     &  -l_3   &  l_2 
  \end{bmatrix} 
   \begin{bmatrix}
     b_1    \\
     b_2    \\ 
     b_3    
   \end{bmatrix}
   = \mathbf{0}.   
\end{equation}
In the above
\begin{align} \label{l2l3l4}
& l_2 = s'_{55} p^2 - 2 s'_{45}p + s'_{44}, \nonumber \\
& l_3 = s'_{15} p^3 - (s'_{14} + s'_{56})p^2 + (s'_{25} + s'_{46})p - s'_{24}, \\
& l_4 = s'_{11} p^4 - 2s'_{16} p^3 + (s'_{66} + 2s'_{12})p^2 - 2s'_{26}p + s'_{22}.
\nonumber
\end{align}
From \eqref{eigenStatics} the eigenvalues $p$ are computed from the sextic equation
\begin{equation} \label{sexticStatics}
l_2 l_4 - l_3 l_3 =0,
\end{equation}
originally given by Lekhnitskii \cite{Lekh63}.  The vector $\mathbf{a}$ is \cite{Ting99}
\begin{equation} \label{aStatics}
 \mathbf{a}=
    \begin{bmatrix}
     g_1        &  -h_1         \\
    p^{-1}g_2  &   -p^{-1}h_2   \\ 
     g_5        &   h_5 
  \end{bmatrix} 
   \begin{bmatrix}
     b_2    \\
     b_3    
   \end{bmatrix}, 
\end{equation}
in which
\begin{equation} \label{g_alpha-h_alpha}
g_\alpha = s'_{\alpha 1} p^2 - s'_{\alpha 6} p + s'_{\alpha 2},
\quad
h_\alpha = s'_{\alpha 5} p - s'_{\alpha 4}.
\end{equation}
We have thus the eigenvalues $p$ and the eigenvectors $\mathbf{a}$ and $\mathbf{b}$  all in terms of $s'_{\alpha\beta}$.

When the material is incompressible, $s'_{2 \beta} = - s'_{1 \beta}$ and the $l_3, l_4$  in \eqref{l2l3l4} simplify to
\begin{align} \label{l3l4Incomp}
& l_3 = (s'_{15} p - s'_{14})(p^2 -1) - s'_{56}p^2 + s'_{46}p,
\nonumber  \\
& l_4 = s'_{11}(p^2 -1)^2 - 2s'_{16} p(p^2 -1) + s'_{66}p^2.
\end{align}
Also, \eqref{g_alpha-h_alpha} gives
\begin{equation} 
g_2 = -g_1,
\quad
h_2= -h_1.
\end{equation}
Equation \eqref{aStatics} can then be written as
\begin{equation} \label{aStaticsIncomp}
 \mathbf{a}=
    \begin{bmatrix}
     g_1        &  -h_1         \\
    -p^{-1}g_1  &   p^{-1}h_1   \\ 
     g_5        &   h_5 
  \end{bmatrix} 
   \begin{bmatrix}
     b_2   \\
     b_3    
   \end{bmatrix}.   
\end{equation}
The $a_1, a_2$  computed from \eqref{aStaticsIncomp} indeed satisfy the identity \eqref{a1a2b1b2}$_1$.
    In the next section we study the special case of incompressible monoclinic materials with the symmetry plane at $x_3=0$.

\section{Monoclinic materials with the symmetry plane at $x_3=0$}

When the material is monoclinic with the symmetry plane at $x_3=0$, $l_3$ vanishes identically so that the sextic equation \eqref{sexticStatics} leads to $l_2=0$ or $l_4=0$.  If the material is incompressible, $l_4$ is given by \eqref{l3l4Incomp} and we have
\begin{equation} \label{p}
(p - p^{-1})^2 - 2 \alpha (p - p^{-1}) + \beta =0,
\end{equation}
where
\begin{equation}
\alpha = s'_{16}/s'_{11}, \quad \beta =  s'_{66}/s'_{11}.
\end{equation}
Since $p_1, p_2$ are the roots of \eqref{p} with a positive imaginary part, \eqref{p} gives
\begin{equation} \label{roots}
p - p^{-1} = \alpha + i \gamma,
\end{equation}
in which
\begin{equation}
\gamma = \sqrt{\beta - \alpha^2} = \sqrt{s'(1,6)}/s'_{11}.
\end{equation}
Equation \eqref{roots} tells us that
\begin{equation} \label{p1p2}
p_1 + p_2 = \alpha + i \gamma, \quad p_1 p_2 = -1.
\end{equation}
We also obtain an explicit expression of $p_1, p_2$ as
\begin{equation} 
p_1, p_2 = \frac{\alpha + i \gamma}{2} \pm \sqrt{(\frac{\alpha + i \gamma}{2})^2 +1}.
\end{equation}

The three Barnett-Lothe \cite{BaLo73} tensors $\mathbf{S}, \mathbf{H}, \mathbf{L}$ appear often in the solutions to anisotropic elasticity problems.  They are real.  Explicit expressions of $\mathbf{S}, \mathbf{H}, \mathbf{L}$ for monoclinic materials with the symmetry plane at $x_3=0$ have been presented in \cite[p.174]{Ting96}.  Specializing to incompressible materials using \eqref{p1p2}leads to
\begin{align}
\label{SHL}
& \mathbf{S} = \mathbf{0},
\quad
\mathbf{H} = \mathbf{L}^{-1} = \text{Diag} [\gamma s'_{11}, \gamma s'_{11}, 1/\mu],\\
& \mu = [s'(4,5)]^\halft.
\end{align}
The quantity $\mu$ is the shear modulus when the material is isotropic.  The structure of $\mathbf{S}, \mathbf{H}, \mathbf{L}$ in \eqref{SHL} provides the following interesting results in elastostatics for incompressible materials.

The order of the stress singularity at an interfacial crack tip in a bimaterial consisting of two dissimilar materials bonded together is not a complex number when $\mathbf{S} \mathbf{L}^{-1}$ in the two materials are identical.  In this case, the physically unrealistic interpenetration of the crack surface displacement does not occur (see, for example, \cite[p.144]{Ting96}).  For a bimaterial for which both materials are incompressible, $\mathbf{S} = \mathbf{0}$ according to \eqref{SHL}.  Hence $\mathbf{S} \mathbf{L}^{-1}$ vanishes in both materials.  Therefore there is no interpenetration of the crack surface when the material is incompressible and monoclinic with the symmetry plane at $x_3=0$.

The inplane displacement and the antiplane displacement for a monoclinic material with the symmetry plane at $x_3=0$ are uncoupled \cite{Stro62}.  We can therefore consider the inplane and antiplane deformations separately.  Consider the inplane deformation.  The Barnett-Lothe tensors now require only the $2 \times 2$ matrix located at the top left corner of $\mathbf{S}, \mathbf{H}, \mathbf{L}$.  From \eqref{SHL} we have
\begin{equation} \label{SHLincomp}
\mathbf{S} = \mathbf{0},
\quad
\mathbf{H} = \mathbf{L}^{-1} = \gamma s'_{11} \mathbf{I},
\end{equation}
where $ \mathbf{I}$ is the $2 \times 2$ identity matrix. Consider now an infinite monoclinic material subject to a line of concentrated force $\mathbf{f}$ and a line of dislocation with Burgers vector $ \widehat{\mathbf{b}}$ applied along the $x_3$-axis.  The strain energy in the annual region bounded by the two radii $r_2>r_1$ can be shown to be 
\begin{equation}\label{strainEnergyCompr}
 \frac{1}{4 \pi} \ln (\frac{r_2}{r_1}) ( \mathbf{f}^T \mathbf{Hf} 
  + \widehat{\mathbf{b}}^T \mathbf{L} \widehat{\mathbf{b}}),
\end{equation}
for a compressible material \cite[p.249]{Ting96}.  When the material is incompressible and when the vectors $\mathbf{f}$ and $ \widehat{\mathbf{b}}$ lie on the $x_3=0$ plane, use of \eqref{SHLincomp} in \eqref{strainEnergyCompr} yields
\begin{equation}
 \frac{1}{4 \pi} \ln (\frac{r_2}{r_1}) [ \gamma s'_{11} |\mathbf{f}| + (\gamma s'_{11})^{-1} |\widehat{\mathbf{b}}|].
\end{equation}
This strain energy depends only on the magnitudes, not the directions, of the vectors  $\mathbf{f}$ and $ \widehat{\mathbf{b}}$.
    
Consider next a half-space with a traction-free boundary surface subject to a line dislocation with Burgers vector $ \widehat{\mathbf{b}}$  in the half-space \cite[pp. 264-265]{Ting96}.  When the material is incompressible it can be shown that, by virtue of \eqref{SHLincomp}, the image force that is attracted to the free-surface depends on the magnitude, not on the direction, of the Burgers vector $ \widehat{\mathbf{b}}$.  Likewise, if the boundary surface is a rigid surface \cite{TiBa93}, the image force that is repelled by the rigid surface depends on the magnitude, not on the direction, of the Burgers vector $ \widehat{\mathbf{b}}$.  Moreover, the magnitude of the repel force is identical to the attracted force when the boundary is a free-surface.

The same result applies to a line dislocation in a bimaterial that consists of two dissimilar materials bonded together \cite[p. 286]{Ting96}.  When the material is incompressible, the image force that is attracted to or repelled by the interface depends on the magnitude, not on the direction, of the Burgers vector.

Clearly, other interesting physical phenomena can be cited when the material is incompressible and monoclinic with the symmetry plane at $x_3=0$.




\end{document}